# Correlation between magnetism and magnetocaloric effect in $RCo_2$-based Laves phase compounds

Niraj K. Singh, K. G. Suresh*

Indian Institute of Technology Bombay, Mumbai- 400076, India

A. K. Nigam and S. K. Malik

Tata Institute of Fundamental Research, Homi Bhabha Road, Mumbai- 400005, India

A. A. Coelho and S. Gama

Instituto de Física "Gleb Wataghin," Universidade Estadual de Campinas-UNICAMP,

C.P. 6165, Campinas 13 083 970, SP, Brazil

*Corresponding author (email: suresh@phy.iitb.ac.in)

# ABSTRACT

By virtue of the itinerant electron metamagnetism (IEM), the RCo$_2$ compounds with R=Er, Ho and Dy are found to show first order magnetic transition at their ordering temperatures. The inherent instability of Co sublattice magnetism is responsible for the occurrence of IEM, which leads to interesting magnetic and related properties. The systematic studies of the variations in the magnetic and magnetocaloric properties of the RCo$_2$-based compounds show that the magnetovolume effect plays a decisive role in determining the nature of magnetic transitions and hence the magnetocaloric effect (MCE) in these compound. It is found that the spin fluctuations arising due to the magnetovolume effect reduce the strength of IEM in these compounds, which subsequently lead to a reduction in the MCE. Most of the substitutions at the Co site are found to result in a positive magnetovolume effect, leading to an initial increase in the ordering temperature. Application of pressure, on the other hand, causes a reduction in the ordering temperature due to the negative magnetovolume effect. A comparative study of the magnetic and magnetocaloric properties of RCo$_2$ compounds under various substitutions and applied pressure is presented. Analysis of the magnetization data using the Landau theory has shown that there is a strong correlation between the Landau coefficient (*B*) and the MCE. The variations seen in the order of magnetic transition and the MCE values seem to support the recent model proposed by Khmelevskyi and Mohn for the occurrence of IEM in RCo$_2$ compounds.

## 1. INTRODUCTION

Twenty first century saw the emergence of magnetism and magnetic materials as major fields of activity which have got profound influence in the industry and day-today life of human beings. Magnetic materials play an integral part in applications such as permanent magnets, memory devices, transducers etc [1]. A relatively new entrant to this list is the magnetic refrigeration. Magnetic materials are the active materials for a magnetic refrigerator [2, 3]. The use of a magnetic material as a refrigerant in a magnetic refrigerator relies on its magnetocaloric behavior. Magnetocaloric effect (MCE) is defined as the cooling or heating of a magnetic material when it is subjected to a varying magnetic field. This magneto-thermal phenomenon was first discovered by Warburg in 1881 [4]. Warburg found that iron got heated up when placed in a magnetic field and when the magnetic field was removed the iron sample cooled down. The origin of MCE was explained independently by Debye and Giauque [5, 6]. They also suggested the first practical use of the MCE: the adiabatic demagnetization used to reach temperatures lower than that of liquid helium, which had been the lowest achievable experimental temperature. Nowadays, there is a great deal of interest in using the magnetic refrigeration as an alternative refrigeration technology, from room temperature to the cryogenic temperature regime [2, 7]. The magnetic refrigeration offers the prospect of an energy-efficient and environment- friendly alternative to the commonly used vapor-cycle refrigeration technology in use today. Furthermore, this technology is purely solid state based and has the advantage that the engineering design remains unaltered even when the refrigerant is changed to suit different operating temperatures. Efforts are on to fabricate magnetic refrigerators for house-hold, industrial and technological applications.

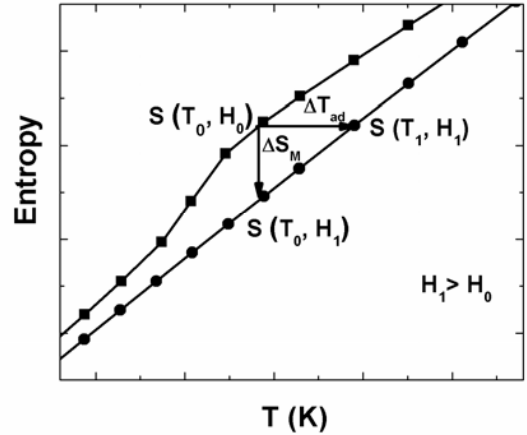

Fig. 1 Schematic S-T plots of a ferromagnetic material illustrating the existence of magnetocaloric effect. The solid lines represent the total entropy in two different magnetic fields: $H_0$ and $H_1$ ($H_1 > H_0$). The horizontal and the vertical arrows show adiabatic temperature change ($\Delta T_{ad}$) and the isothermal magnetic entropy change ($\Delta S_M$), respectively, when the magnetic field is changed from $H_0$ to $H_1$.

## 2. MAGNETOCALORIC EFFECT AND MAGNETIC REFRIGERATON

The main challenges in the realization of a practical refrigerator are (i) the availability of bulk amounts of giant magnetocaloric materials (ii) proper magnetic field design, preferably with a permanent magnet array and (iii) ingenious design. Since most of the materials developed so far show giant magnetocaloric effect (GMCE) only at fields as high as 50 kOe, there is an urgent need to look for more potential materials so that the same MCE could be obtained with lower fields. In this context, materials which show first order magnetic transitions, metamagnetic transitions or magneto-structural transitions are being probed.

The MCE is intrinsic to all magnetic materials and is induced via the coupling of magnetic sublattice with the applied magnetic field and is measured as adiabatic temperature change ($\Delta T_{ad}$) or isothermal entropy change ($\Delta S_M$). The isothermal magnetization of a paramagnet or a soft ferromagnet reduces the entropy and, in a reversible process, demagnetization (which is similar to the expansion of a gas) restores the zero-field magnetic entropy of the system. Since in an adiabatic process the total entropy of a material remains constant, the adiabatic demagnetization of a ferromagnetic material leads to reduction in the temperature.

The value of the total entropy (S) of a ferromagnet at constant pressure depends on both the magnetic field (H) and temperature (T) and the contribution to it arises from the lattice ($S_{lat}$), electronic ($S_{el}$) and the magnetic ($S_M$) entropies,

$$S(T,H) = S_M(T,H) + S_L(T) + S_{el}(T) \tag{1}$$

Fig. 1 schematically shows the S *vs* T plots of a ferromagnet in two different fields $H_0$ and $H_1$ ($H_1 > H_0$). In order to understand the thermodynamics of MCE, two relevant processes are shown in the Fig. 1:

(i) When the magnetic field is applied adiabatically in a reversible process, the magnetic entropy decreases, but as the total entropy does not change, i.e.,

$$S(T_0, H_0) = S(T_1, H_1) \tag{2}$$

it leads to an increase in the temperature of the material. The adiabatic temperature change ($\Delta T_{ad}$), i.e. the difference between the initial temperature $T_0$ and the final temperature $T_1$, can be visualized as the isentropic difference between the corresponding $S(T,H)$ functions and it is a measure of the MCE of the material.

(ii) When the magnetic field is applied isothermally (i.e. keeping T constant), the total entropy decreases due to the increase in the magnetic order. The isothermal magnetic entropy change ($\Delta S_M$) in the process is defined as

$$\Delta S_M = S(T_0, H_1) - S(T_0, H_0) \tag{3}$$

Therefore, it can be seen that $\Delta T_{ad}$ and $\Delta S_M$ represent the two quantities which are characteristic of MCE. It is obvious from Fig. 1 that both $\Delta T_{ad}$ and $\Delta S_M$ are functions of the initial temperature $T_0$ and the magnetic field change $\Delta H = H_1 - H_0$. Furthermore, it is easy to see from the figure that if raising the magnetic field increases the magnetic order, $\Delta T_{ad}$ is positive and the magnetic material heats up ($\Delta S_M$ is negative). The signs of $\Delta T_{ad}$ and $\Delta S_M$ are correspondingly reversed when the magnetic field is reduced.

The MCE parameters viz. $\Delta T_{ad}$ and $\Delta S_M$ are correlated with the magnetization (M), the magnetic field strength, the heat capacity (C) and the absolute temperature by one of the fundamental Maxwell's relations [2].

$$\left(\frac{\partial S(T,H)}{\partial H}\right)_T = \left(\frac{\partial M(T,H)}{\partial T}\right)_H \tag{4}$$

For an isothermal-isobaric process the integration of the above equation yields

$$\Delta S_M(T, \Delta H) = \int_{H_0}^{H_1} \left(\frac{\partial M(T,H)}{\partial T}\right)_H dH. \tag{5}$$

Equation 5 indicates that the magnetic entropy change is proportional to the temperature derivative of magnetization at constant field. By combining the equation 4 with the corresponding TdS equation, it can be shown that the infinitesimal adiabatic temperature rise for the reversible adiabatic-isobaric process is equal to

$$dT = -\left(\frac{T}{C(T,H)}\right)_H \left(\frac{\partial M(T,H)}{\partial T}\right)_H dH. \qquad (6)$$

Hence, it may be noted from the above equation that the adiabatic temperature rise is directly proportional to the absolute temperature, the temperature derivative of magnetization and the magnetic field change. However, it is inversely proportional to the heat capacity. The integration of equation 6 gives the value of MCE as

$$\Delta T_{ad}(T,\Delta H) = -\int_{H_0}^{H_1} \left(\frac{T}{C(T,H)}\right)_H \left(\frac{\partial M(T,H)}{\partial T}\right)_H dH. \qquad (7)$$

A careful examination of the equations 4 to 7 reveals the following points:

(i) For a ferromagnet the $|(\partial M/\partial T)_H|$ is largest at the ordering temperature ($T_C$), therefore, $|\Delta S_M(T)_{\Delta H}|$ should peak at $T_C$.

(ii) Using equations 6 and 7, Tishin *et al.* have shown that, for a ferromagnet, in the limit of $\Delta H$ tending to zero, $\Delta T_{ad}$ also peaks near the $T_C$ [8]. Furthermore, the behavior of the $\Delta T_{ad}$ and $|\Delta S_M(T)_{\Delta H}|$ should be similar and they will gradually be reduced on both sides of the $T_C$.

(iii) For the same value $|\Delta S_M(T)_{\Delta H}|$, the $\Delta T_{ad}$ will be larger at higher absolute temperature, and also when the total heat capacity of the solid is lower. Furthermore, at elevated temperatures, due to large value of the total heat capacity, considerable $\Delta T_{ad}$ can be observed only when $|(\partial M/\partial T)_H|$ becomes significant.

From the above discussion, it is clear that the MCE is significant only close to a magnetic transition temperature. As a consequence, in general ferro-para magnetic transitions will show the maximum entropy change. Therefore, the first and foremost attempt in the design of a magnetic refrigerator is to identify a suitable magnetic material which shows a magnetic transition near the operating temperature of the refrigerator. Since the MCE depends on the sharpness of the transition, materials with first order magnetic transitions are preferred. Another alternative is to use materials which undergo magnetic field-induced structural transitions. The entropy change associated with the structural changes would enhance the magnetic entropy change and hence a large MCE is expected. Therefore, from the point of view of refrigeration applications, one looks for materials which exhibit giant magnetocaloric effect. However, when put into the application, there will always be a drift in the operating temperature and therefore, the refrigerant system should contribute considerable MCE over a span of temperature around the operating temperature. This can be achieved by making a composite magnetic material which contains many materials whose magnetic transition temperatures are spread over a span around the operating temperature. Other possibility is to have a material which shows close-by multiple magnetic transitions. In both these cases, a 'table-like' (flat) MCE vs. T plot is expected. However, due to the lack of materials with sharp multiple transitions, emphasis is given to the first choice, i.e., to tune the magnetic transition of the parent system with the help of substitutions so that a graded magnetic material is produced. But, substitutions generally reduce the MCE and therefore, a compromise has to be reached so as to get a reasonable MCE spread over a range of temperature or a 'table-like' MCE.

## 2. 1     MEASUREMENT OF MCE USING INDIRECT METHODS

Unlike direct measurement, which usually yields only the adiabatic temperature change, indirect method using the heat capacity data at different applied fields allow the calculation of both $\Delta T_{ad}(T, \Delta H)$ and $\Delta S_M(T, \Delta H)$. On the other hand, the calculation using magnetization measurements yields only $\Delta S_M(T, \Delta H)$. In the latter case, magnetization must be measured as a function of T and H. This allows to obtain $\Delta S_M(T, \Delta H)$ by numerical integration using equation. 5, and it is a very useful tool in the rapid search for potential magnetic refrigerant materials [2, 9, 10]. The accuracy of $\Delta S_M(T,\Delta H)$ calculated from magnetization data depends on the accuracy of the measurements of the magnetic moment, T and H. It is also affected by the fact that the exact differentials in equation 5 (dM, dH and dT) are replaced by the measured variations (dM, dT and dH). Taking into account all these effects, the error in the value of $\Delta S_M(T, \Delta H)$ lies within the range of 3-10% [2, 10]. The measurement of the heat capacity as a function of temperature in constant magnetic fields and pressure, $C(T)_{P,H}$, provides the most complete characterization of MCE in magnetic materials. The entropy of a solid can be calculated from the heat capacity as:

$$S(T)_{H_0} = \int_0^T \frac{C(T)_{H_0}}{T} dT + S_{H_0} \qquad (8)$$

$$S(T)_{H_1} = \int_0^T \frac{C(T)_{H_1}}{T} dT + S_{0,H_1} \qquad (9)$$

where $S_{H_0}$ and $S_{H_1}$ are the zero temperature entropies. In a condensed system $S_{H_0} = S_{H_1}$ [11]. Hence, if $S(T)_H$ is known, both $\Delta T_{ad}(T, \Delta H)$ and $\Delta S_M(T, \Delta H)$ can be obtained [10] with he help of the following equations (see Fig. 1):

$$\Delta T_{ad}(T)_{\Delta H} \cong [T(S)_{H_1} - T(S)_{H_0}]_S \qquad (10)$$

$$\Delta S_M(T)_{\Delta H} = S(T)_{H_1} - S(T)_{H_0} \qquad (11)$$

On the basis of the magnetocaloric properties of a number of compounds, Pecharsky and Gschneidner have shown that the determination of MCE using the indirect methods are reliable and compares well with those determined from the direct measurement [12].

## 2.2     POTENTIAL MAGNETOCALORIC MATERIALS FOR VARIOUS TEMPERATURE REGIMES

The magnetocaloric effect can be exploited for cooling applications in various temperature regimes, the oldest being the adiabatic demagnetization which was used to achieve the temperatures below 1 K with the help of paramagnetic salts such as Gd$_2$(SO$_4$)$_3$·8H$_2$O [13]. Apart from the paramagnetic salts, the paramagnetic alloys like PrNi$_5$ has also been used in the adiabatic demagnetization devices [14]. Recent studies on the paramagnetic garnets and their nanocomposites have shown that these materials are also potential refrigerants in the temperatures regime below 10 K [15]. Due to the low value of $\left|(\partial M/\partial T)_H\right|$, the paramagnetic salts yield small $\Delta T_{ad}$ and hence are not suitable refrigerants for the temperatures above 10 K. For this temperature range, magnetically ordered materials have to be used. Because of the presence of the magnetic correlations, the magnetically ordered materials undergo an order-disorder transition in a narrow temperature intervals and therefore yield large value of $\left|(\partial M/\partial T)_H\right|$, which in turn leads to considerable MCE. It has been reported that pure rare earth metals such as Pr, Nd, Er and Tb exhibit considerable MCE below 60 K [2]. However, the investigations regarding the MCE of various rare earth alloys and intermetallic compounds suggest that they are better refrigerants than the pure rare earths. Among them, the best for the range of 10-80 K are RAl$_2$, Dy$_{0.5}$Ho$_{0.5}$, Dy$_x$Er$_{1-x}$ [0<x<1], RNiAl and RNi$_2$ compounds [2, 16-18]. For temperatures above 80 K, the promising refrigerants are Dy metal and Gd$_5$(Si,Ge)$_4$ alloys [19]. However, near room temperature, the prototype material is pure Gd [2]. Most of the intermetallic compounds which order near room temperature show significantly lower MCE as compared to that of Gd metal [20]. The only system in which the MCE is equal to that of Gd is Gd$_5$(Si,Ge)$_4$. Apart from the rare earth based intermetallic compounds, a few transition metal based compounds are also known to exhibit considerable MCE near room temperature. The best among them are MnFeP$_{1-x}$As$_x$, Mn(As,Sb) [3, 21, 22].

## 3   MAGNETISM OF RCo$_2$ COMPOUNDS AND ITS IMPORTANCE IN MAGNETOCALORIC EFFECT

Rare earth (R) – transition metal (TM) intermetallics formed between different rare earths and TM=Fe, Co, Ni crystallize in the cubic structure with MgCu$_2$-type structure [23]. They also form with nonmagnetic elements such as Al, Ga, Si, Ge etc. The magnetic ordering temperatures of RFe$_2$ compounds being much above the room temperature, from the point of view of magnetic refrigeration application, they are not considered. On the other hand the compounds based on Co and Ni have their ordering temperatures below room temperature and has attracted several studies. The tunability of the magnetic ordering temperature of these compounds over a wide span with the help of substitutions at the Co/Ni site, without changes in the crystal structure, has prompted many researchers to carry out detailed magnetic and other related investigations [24-30].

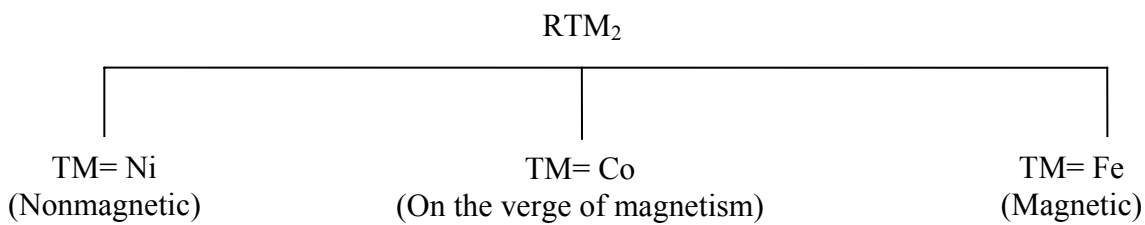

Fig. 2 Schematic diagram showing the magnetic state of TM sublattices in RTM$_2$ compounds.

Though RTM$_2$ [TM=Fe, Co and Ni] compounds crystallize in the same structure, there are considerable differences in their magnetic properties, especially with regard to the TM component [31]. This is due to the peculiar band structure of these compounds [23]. The hybridization between the transition metal 3d band and the rare earth 5d (4d) band gives rise to interesting magnetic properties in R-TM intermetallic compounds. In R-TM intermetallics, for a particular R atom, the change in the number of 3d electrons across the TM series leads to considerable effect on the density of state s at the Fermi level [$N(E_F)$] and plays a decisive role in determining the magnetic properties. Within the RTM$_2$ series, RFe$_2$ compounds posses magnetic transition metal sublattice with a magnetic moment of 1.5$\mu_B$ [32]. The complete replacement of Fe by Co gives rise to one extra electron per Co atom to the 3d band. In the rigid band picture, the addition of the extra electrons leads to the filling of the 3d band, which in turn alters the $N(E_F)$ and gives rise to interesting magnetic properties in the RCo$_2$ compounds. In this series of compounds the magnetic state of Co strongly depends on the alloying rare earth and the moment varies between 0 and 1 $\mu_B$ [32]. The variation within the series indicates the inherent instability of the magnetic state of Co sublattice. In the RNi$_2$ compounds two extra 3d electrons, as compared to RFe$_2$, per Ni atom are added. This further addition of the extra electron, as compared to that of RCo$_2$ compounds, drives the Fermi level to a region of low density of states and therefore leads to a permanently nonmagnetic state of the 3d sublattice in the RNi$_2$ compounds. Fig. 2 shows the schematic diagram showing the TM magnetic state in RTM$_2$ compounds.

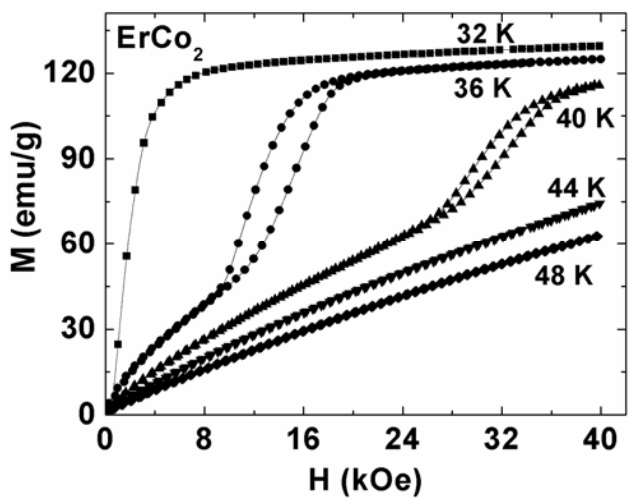

Fig. 3 M vs. H isotherms showing the itinerant electron metamagnetism in ErCo$_2$

In RCo$_2$ compounds, the $T_C$ values vary from ~20 K to ~ 400 K. Unlike the case of the YFe$_2$, the YCo$_2$ does not exhibit any magnetic ordering and behaves like an exchange enhanced Pauli paramagnet [23]. The instability of the magnetic state associated with the Co sublattice mentioned earlier in some of the RCo$_2$ compounds leads to metamagnetic transition. The creation of the Co moments by the R molecular field as these samples are cooled through their ordering temperatures is responsible for the metamagnetic transition, which is termed as itinerant electron metamagnetism (IEM). Fig.3 shows the M-H isotherms illustrating the IEM at temperatures just above $T_C$ in $ErCo_2$. The IEM in these compounds leads to first order transition (FOT) at their ordering temperature and many interesting observations regarding their structural, magnetic, thermal and transport behavior. Among the $RCo_2$ compounds, $ErCo_2$, $HoCo_2$ and $DyCo_2$ are found to show FOT while all others show a second order transition (SOT) at their ordering temperature. Based on a systematic analysis of the magnetic and electrical resistivity data on $(Er,Y)Co_2$, Hauser et al. have indeed shown that dilution of Er by Y leads to the disappearance of IEM [33]. The IEM in these compounds may also be induced by an external magnetic field in a limited temperature range above $T_C$ [23]. In most of the magnetic materials, the transitions are of second order nature and hence IEM systems have received a lot of attention for decades owing to their interest in fundamental physics. However, of late they have become natural probes in the search for materials with large magnetocaloric effect as well as magnetoresistance, by virtue of the first order nature of their magnetic transitions. Other IEM systems such as $La(Fe,Si)_{13}$-based systems have also attracted considerable attention from the point of view of MCE recently [34].

## 3.1    ITINERANT ELECTRON METAMAGNETISM IN $RCo_2$ COMPOUNDS

It has been observed that some $RCo_2$ compounds [R=Er, Ho and Dy] undergo IEM in which the Co sublattice changes from a nonmagnetic state to a magnetic state when the field acting on the 3d band is larger than a critical value $H_C$ [23, 33, 35]. This field-induced transition in the 3d sublattice is known as itinerant electron metamagnetism and was first theoretically predicted by Wohlfarth and Rhodes in 1962 [36]. Using the Landau expansion of the magnetic free energy of the d-electrons as

$$F = \frac{A}{2}M^2 + \frac{B}{4}M^4 + \frac{C}{6}M^6 + \ldots\ldots - MH \quad (12)$$

The theory led to the following expressions for the first lower order coefficients

$$A = \frac{1}{4N(E_F)R} \quad (13)$$

$$B = -\frac{1}{64N^3(E_F)}\left[\frac{N''(E_F)}{3N(E_F)} - \left\{\frac{N'(E_F)}{N(E_F)}\right\}^2\right] \quad (14)$$

where $R = [1 - IN(E_F)]^{-1}$ is the Stoner enhancement factor. $N(E_F)$, $N'(E_F)$ and $N''(E_F)$ are the density of states and its first and second derivatives at Fermi level. IEM occurs when the magnetization dependence of free energy has variation similar to that shown in Fig. 4a in zero field and it changes to the lower curve of this figure when field is increased to higher fields ($H_1<H_C<H_2$) [36]. The corresponding magnetization curve is shown in Fig. 4b. It has been reported that such magnetization dependence of free energy and hence the metamagnetic transition in the field dependence of magnetization arises when coefficient of $M^4$ is positive and some of higher order terms in the equation 12 give negative contribution. Therefore, it may be noticed from equations 13 & 14 that IEM can occur when: (1) *A* is weakly positive. i. e. when the Stoner criterion for the onset of ferromagnetism is almost satisfied, and (2) *B* is negative, which implies another minimum for a non-zero value of *M*. The later condition requires $N''(E_F)$ to be large enough, which in turn means that the density of states at Fermi level should have strong positive curvature. On theoretical grounds, Wohlfarth and Rhodes showed that a maximum in the temperature dependence of magnetic susceptibility is expected due to the large positive curvature of the density of states and is characteristic of an IEM system [36]. Such a temperature dependence of susceptibility has indeed been observed in a number of compounds which exhibit IEM [37].

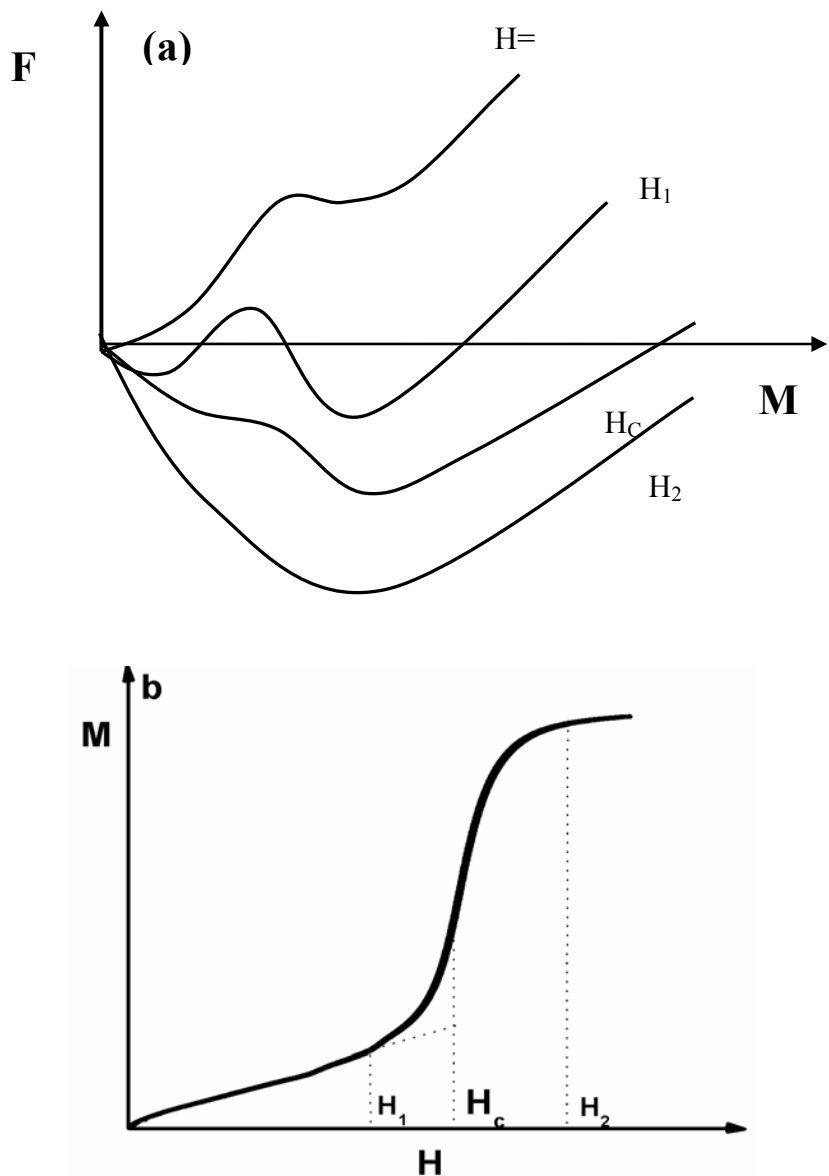

Fig. 4. (a) Magnetization dependence of free energy of a metamagnetic system at different fields ($H_1<H_C<H_2$); (b) Field dependence of magnetization of a metamagnetic system.

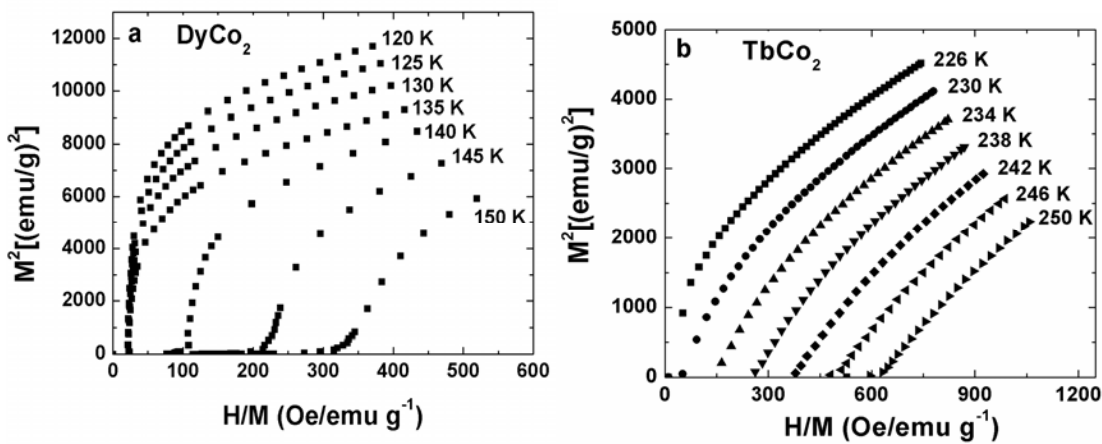

Fig. 5 Arrott's plots of DyCo$_2$ and TbCo$_2$ [Fig. 5a taken from Ref. 35.]

It was mentioned above that materials with negative $M^4$ coefficient in equation 12 exhibit IEM and FOT. In general, the IEM is accompanied with the hysteresis effect (see Fig. 3). However there are reports in literature that, even though some materials posses negative $M^4$ coefficient, they do not distinctly exhibit IEM and hence the hysteretic effect. In such cases, the presence of IEM and FOT are diagnosed with the help of Arrott's plots. It has been established that the materials with negative $M^4$ coefficient in the Landau free energy expression posses S-shaped Arrott's plots [38]. To emphasize on the above point, the Arrott's plots for DyCo$_2$ and TbCo$_2$ are shown in Figs. 5a &b, respectively. It can be clearly seen from these figures that the near $T_C$ Arrott's plots of DyCo$_2$, which exhibits FOT, are S-shaped whereas those corresponding to TbCo$_2$, which is known to possess SOT, are linear.

### 3.2 ROLE OF MAGNETOVOLUME EFFECT IN RCo$_2$-BASED COMPOUNDS

Considerable effort has been put to understand the mystery behind the changeover of the order of magnetic transition from SOT to FOT, as the rare earth is changed in RCo$_2$ compounds. The first true attempt towards understanding the nature of transition in these compounds was made by Bloch *et al.* and later by Inoue and Shimizu [39, 40]. According to these authors, the critical parameter that governs the IEM and therefore FOT in these compounds is the molecular field or the ordering temperature [39, 40]. However, these theories met with only limited success. Recently, Khmelevskyi and Mohn have modified these models by incorporating the contributions from the magnetovolume effect and spin fluctuations [41]. The most important feature in the modified model is the role of lattice parameter in determining the magnetic state of Co and thereby the possibility of IEM and FOT.

Table 1 Magnetic ordering temperature, maximum values of isothermal entropy change and adiabatic temperature change in substituted $RCo_2$ compounds [R=Er, Ho and Dy]. $\Delta H$ is the field change for which the MCE is calculated.

| Compound | $T_C$ (K) | $\Delta H$ (kOe) | $(\Delta S_M)^{max}$ (J Kg$^{-1}$K$^{-1}$) | $(\Delta T_{ad})^{max}$ (K) |
|---|---|---|---|---|
| $ErCo_2$ | 35[42] | 50 | 33[42], ~36[24] | ~9.5[24] |
| $Er(Co_{0.95}Si_{0.05})_2$ | ~52[43] | 60 | 20[43] | ~7.5[43] |
| $Er(Co_{0.925}Al_{0.075})_2$ | 90[44] | | | |
| $Er(Co_{0.97}Ga_{0.03})_2$ | 41[26] | | | |
| $Er(Co_{0.97}Ge_{0.03})_2$ | 47[26] | | | |
| $Er(Co_{0.97}In_{0.03})_2$ | 51[26] | 20 | | 1.3[26] |
| $Er(Co_{0.9}Ni_{0.1})_2$ | 13[24] | 50 | 29.8[24] | 9.5[24] |
| $HoCo_2$ | ~78[28] | 50 | 23[28] | ~7.5[28] |
| $Ho(Co_{0.95}Si_{0.05})_2$ | 100[45] | 60 | ~20 | ~6[45] |
| $Ho(Co_{0.95}Al_{0.05})_2$ | 110[46] | | | |
| $Ho(Co_{0.9}Ga_{0.1})_2$ | 165[47] | | | |
| $Ho(Co_{0.9}Ni_{0.1})_2$ | 40[28] | 50 | 22[28] | 8[28] |
| $DyCo_2$ | 142[27] | 50 | 11.3[27] | 5.4[27] |
| $Dy(Co_{0.925}Si_{0.075})_2$ | 164[27] | 50 | 6.5[27] | 3.6[27] |
| $Dy(Co_{1.96}Al_{0.04})_2$ | 171[29] | 10 | 4.3[29] | |
| $Dy(Co_{1.9}Ge_{0.1})_2$ | 173[29] | 10 | ~1.8[29] | |
| $Dy(Co_{1.9}Ga_{0.1})_2$ | 195[29] | 10 | ~1.9[29] | |
| $Dy(Co_{1.94}Fe_{0.06})_2$ | 242[30] | 10 | 1.6[30] | |

The calculations of Khmelevskyi and Mohn suggest that in $RCo_2$ series, the compounds with lattice parameter more than 7.22 Å possess permanently magnetic Co sublattice and therefore show SOT at $T_C$. On the other hand, if the lattice parameter is less than 7.05 Å, they posses nonmagnetic Co 3d sublattice, which is similar to that in $YCo_2$. In the compounds with the lattice parameters in the range of 7.05 - 7.22 Å, the Co sublattice is nonmagnetic, but moment can be induced with the help of the molecular field of the rare earth or the applied field. This transition is called IEM, which results in FOT in such compounds. Fig. 6 shows the theoretical prediction of Khmelevskyi and Mohn, regarding the order of transition for various $RCo_2$ compounds. It is quite clear from this figure that, except $TbCo_2$ and $TmCo_2$, in all other compounds the nature of magnetic transition is consistent with the value of the lattice parameter. It is of importance to note that, based on the studies of a number of R(Co,Al)$_2$ compounds, Duc *et al.* have suggested a critical lattice parameter of 7.27 Å for the appearance of Co 3d magnetic moment in $RCo_2$ compounds [32]. In view of these observations, there is a great scope for studying the structure-magnetic property correlation in $RCo_2$ compounds under various substitutions, which cause lattice parameter variations. Another point of interest is the study of the magnetism and related properties as a function of applied pressure.

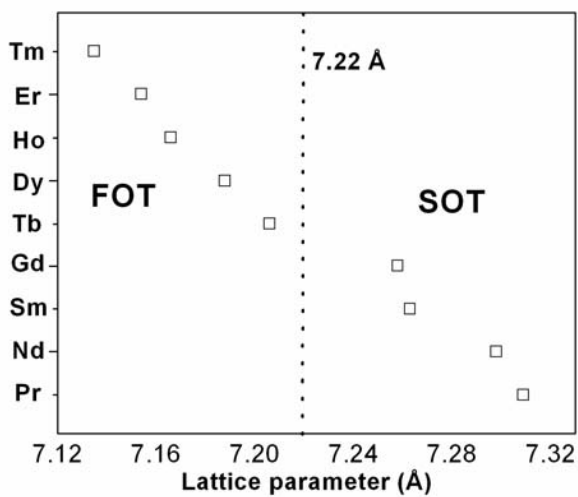

Fig. 6 The predicted order of transition in various $RCo_2$ compounds. [Data taken from Ref. 41.]

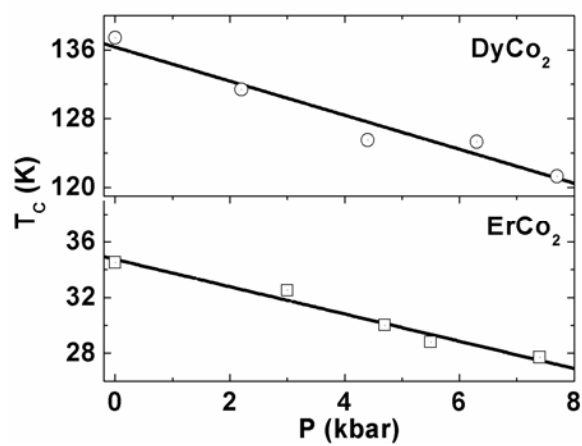

Fig. 7 Pressure dependence of $T_C$ in $ErCo_2$ and $DyCo_2$ compounds. [Data taken Ref. 42.]

There are several reports in the literature which show the strong lattice volume dependence of the magnetic properties in the $RCo_2$ compounds [23, 32, 47, 48]. The role of the lattice parameter in $RCo_2$ compounds has been demonstrated by pressure dependent studies as well [48-50]. Pressure dependence of magnetic properties in $RCo_2$ compounds have been reported by many authors, who have shown that the negative magnetovolume effect produced by the pressure destabilizes the Co magnetic state, thereby reducing the $T_C$ values. Fig 7 shows the pressure dependence of $T_C$ in $ErCo_2$ and $DyCo_2$ compounds.

13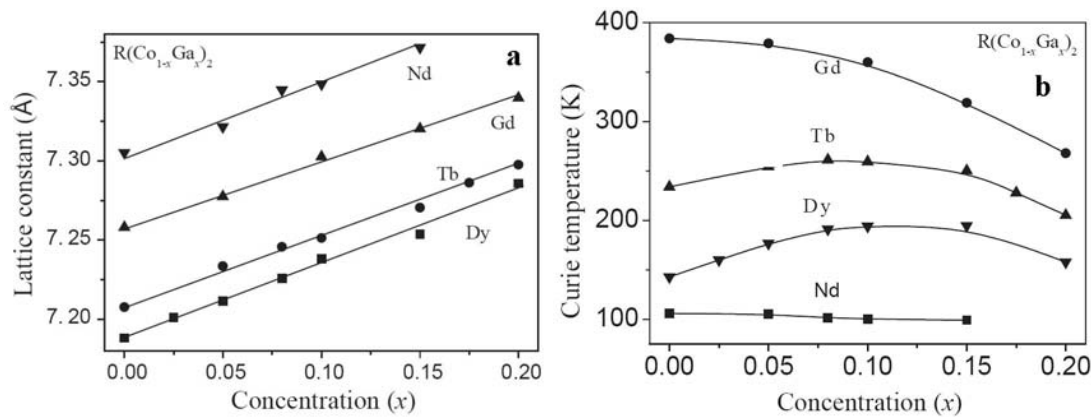

Fig. 8 The concentration dependence of lattice parameter and $T_C$ in $R(Co_{1-x}Ga_x)_2$ compounds. [Figure taken from Ref. 47.]

Effect of substitutions of nonmagnetic elements such as Al, Ga, Ge, In and Si for Co in $RCo_2$ compounds with R=Er, Ho and Dy have been reported by many authors. Table 1 shows the highlights of these studies. As is evident from this table, the ordering temperature, in general, increases with substitutions. As expected, substitution of Fe also causes an increase in $T_C$. The enhancement of $T_C$ due to the substitutions other than Fe mainly arises from the positive magnetovolume effect. The increase in lattice parameter as a result of substitutions (upto a critical concentration) causes local moment formation due to the positive magnetovolume effect. These local moments cause an increase in the R-Co and Co-Co exchange interactions, giving rise to an enhancement of $T_C$ [35]. However, above the critical concentration, the nonmagnetic substitution causes a nearly uniform distribution of Co moments. At this stage, the increasing trend of $T_C$ stops and for further increase in the concentration of the substituents, $T_C$ starts decreasing. Figures 8a & b show the variation of the lattice parameter and the $T_C$ as a function of Ga concentration in $R(Co_{1-x}Ga_x)_2$ compounds, respectively. It can be seen from these figures that the lattice parameters monotonically increase with Ga concentration in all the compounds, but the $T_C$ variation is different for different rare earths. While the compounds based on Dy show an initial increase in $T_C$, the ones with Nd and Gd show a decreasing trend of $T_C$ even at very small concentrations of Ga. The behavior of Tb-based compounds is almost identical to that of Dy compounds. These variations are very much consistent with the predictions of Khmelevskyi and Mohn (see Fig. 6). The prediction that $TbCo_2$ is at the boundary between FOT and SOT is reinforced by the variation of $T_C$ shown in Fig. 8b. Therefore, the examples of nonmagnetic substituted compounds demonstrate the presence of positive magnetovolume effect, in contrast to the negative magnetovolume effect arising from the application of pressure. In the case of Fe substitution, the $T_C$ shows a monotonic increase with Fe concentration, in all the compounds. The increase in $T_C$ seen in the case of Fe substitution mainly arises from the enhanced TM-TM exchange interaction and not from the magnetovolume effect. Han et al have reported that in $Dy(Co,Fe)_2$ systems, the order of transition changes from first order to second order upon Fe substitution. [30].

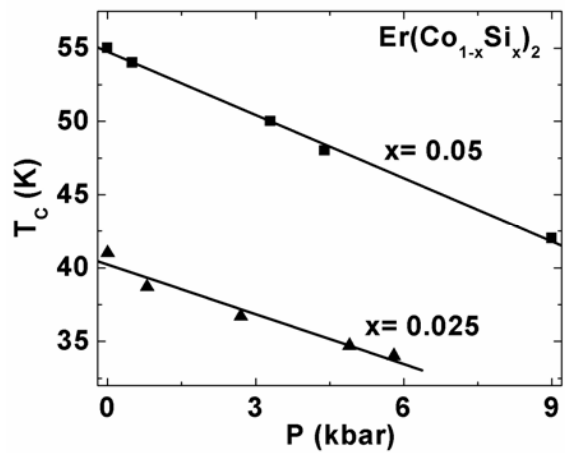

Fig. 9. Variation of $T_C$ as a function of pressure in Er(Co$_{1-x}$Si$_x$)$_2$ compounds. [Data taken from Ref. 42]

In view of the opposing effects of nonmagnetic substitutions and applied pressure on the magnetovolume effect in RCo$_2$ compounds, it is of interest to find out the effect of pressure on these substituted compounds. Fig. 9 shows the effect of pressure on Si substituted ErCo$_2$ compounds. As is evident from this figure, like the parent compounds, the application of pressure causes a reduction in $T_C$ in the Si substituted compounds as well.

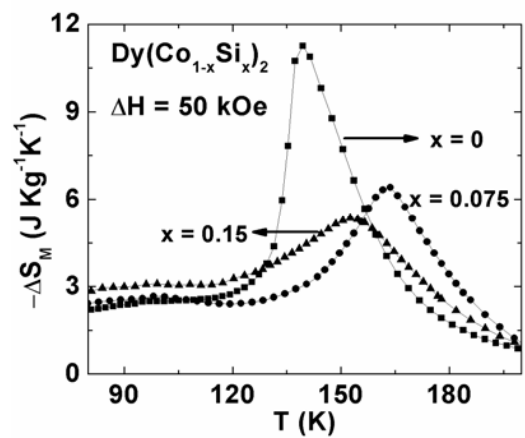

Fig.10 Temperature variation of isothermal magnetic entropy change in Dy(Co$_{1-x}$Si$_x$)$_2$ with x= 0, 0.075 and 0.15] compounds obtained for a field change (ΔH) of 50 kOe. [Data taken from Ref. 27.]

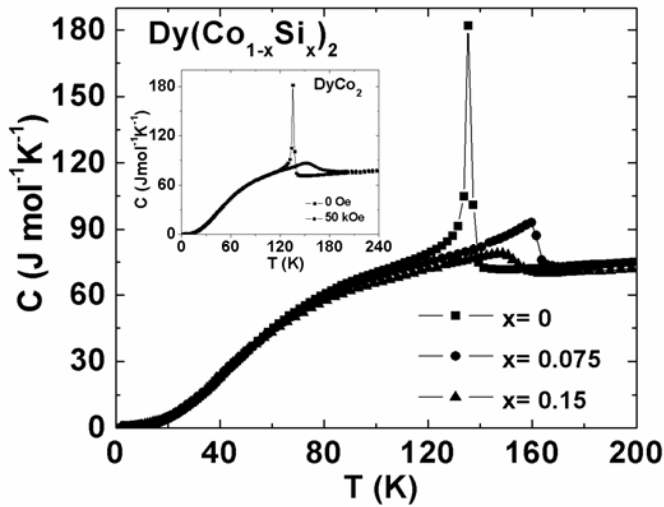

Fig.11 Temperature variation of the zero-field heat capacity (C) of Dy(Co$_{1-x}$Si$_x$)$_2$ compounds [with x=0, 0.75 and 0.15]. The inset shows the C-T data of DyCo$_2$ obtained in zero field and 50 kOe. [Data partly taken from Ref. 27]

## 4  MAGNETOCALORIC EFFECT IN RCo$_2$-BASED COMPOUNDS

Magnetocaloric effect in terms of isothermal magnetic entropy change as well as adiabatic temperature change on RCo$_2$ compounds with R=Er, Ho and Dy have been measured by many authors and the important results are summarized in Table 1. The values obtained in parent RCo$_2$ compounds with Er, Ho and Dy are quite high as compared to many other materials [51]. It is of interest to note that, though the theoretical magnetic entropy i.e. *Rln(2J+1)* value for all these three compounds are approximately equal, the MCE is the highest in ErCo$_2$. As the magnetic entropy tends to saturate around T$_C$ and as the saturation magnetization of these three compounds are very close, the smaller the T$_C$ the faster the magnetic entropy saturation. Accordingly, the maximum entropy change i.e., $\Delta S_M^{max}$ is the highest in the compound with the lowest T$_C$. At this point it is of importance to mention that the $\Delta S_M^{max}$ value in TbCo$_2$ has been found to be about 6 J kg$^{-1}$K$^{-1}$ [25, 52]. Compared to the MCE of the RCo$_2$ compounds with SOT, the MCE seen in TbCo$_2$ is high, which once again justifies the prediction of Khmelevskyi and Mohn (Fig. 6).

In order to find the effect of substitutions on MCE, several studies have been devoted to measure the MCE of substituted RCo$_2$ [R=Er, Ho and Dy] compounds [24-30, 35, 42, 43, 45]. As in the case of other magnetic properties such as ordering temperature, the MCE is also found to be very much affected by substitutions. Fig. 10 shows the MCE variation observed in Si substituted DyCo$_2$. It has been found that the MCE calculated using the C-H-T data as well as the M-H-T data yield almost same results. Fig. 11 shows the typical heat capacity (C) vs. T plot of Dy(Co$_{1-x}$Si$_x$)$_2$ compounds. It can be seen that the onset of magnetic ordering in DyCo$_2$ is characterized by huge jump in the C-T plot, however, the near T$_C$ jump seen in the C-T data of the Si substituted compounds decrease with increasing Si concentration. This observation indicates the change in the order of transition from first order towards second order and corroborates with the magnetization studies [35]. Similar observations have been reported by many other authors on other RCo$_2$-based systems. The inset of Fig. 11 shows the C-T plots of DyCo$_2$ obtained under zero field and in a field of 50 kOe. The variation of MCE under various substitutions is shown in Table 1. As is evident, substitutions, in general, tend to increase the $T_C$ and decrease the MCE values in Er, Ho and Dy compounds. Fig. 12 shows the effect of substitutions on the magnetic entropy change of several RCo$_2$ compounds.

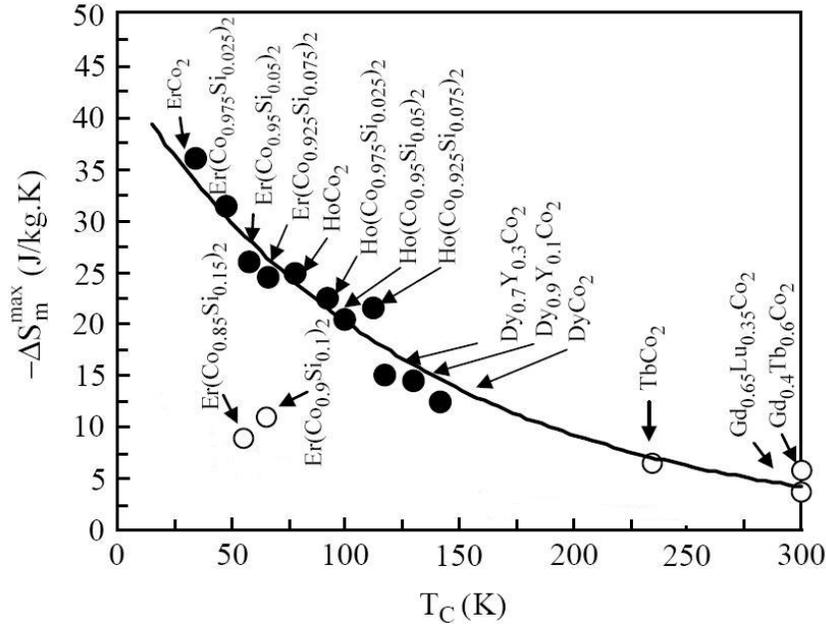

Fig. 12 $\Delta S_M^{max}$ vs. $T_C$ of RCo$_2$-based compounds. The filled circles correspond to the materials which show FOT whereas the open circles correspond to the materials with SOT. [Fig taken from Ref. 25]

Its has been argued by several authors that in RCo$_2$ compounds the substitution of Co by small amounts of magnetic or nonmagnetic elements does not alter the crystalline electric field and therefore the R-moments remain unaltered [28]. Now, since the magnetization tends to saturate around the $T_C$, the resulting variations in the MCE are assumed to be natural consequence of the $T_C$ variation. However, a careful examination of the magnetization and MCE data clearly reveals that apart from the $T_C$ variations, the nature of the transition does play a decisive role in determining the MCE, at least whenever there is an increase in $T_C$. At this juncture it is worth mentioning the report on Dy(Co$_{0.925}$Si$_{0.075}$)$_2$ and Dy(Co$_{0.85}$Si$_{0.15}$)$_2$ compounds, which have the $T_C$ values of 164 K and 154 K, respectively, with maximum $\Delta S_M$ values of 6.5 and 5.3 Jkg$^{-1}$K$^{-1}$ [27]. Based on the data on Er(Co,Si)$_2$ compounds, a similar observation has been reported by Duc *et al.* as well [25]. Therefore, the reduction in MCE due to substitutions at the Co site could be explained as follows. The local magnetic moments arising from the positive magnetovolume effect are not fully exchange coupled and therefore enhances the spin fluctuations. The presence of spin fluctuations decreases the strength of IEM and consequently weakens the first order nature of the transition at $T_C$. Wang et al. have studied the effect of nonmagnetic substitutions such as Al, Si, Ga and Ge for Co in DyCo$_2$ and found that the substitutions cause an increase in T$_C$, but the transition changes from FOT to SOT [29]. Depending on the stability of the Co magnetic state, the concentration (of the substituent) at which the FOT changes to SOT differs from one compound to another. The detrimental role of the spin fluctuations on the strength of IEM and MCE has been reported by Han *et al.* as well [30]. The presence of increased spin fluctuations for small concentrations of nonmagnetic substituents on the properties such as electrical resistivity of RCo$_2$ compounds has also been investigated in the past [53]. It is of interest to note that there exists a correlation between the variations of T$_C$ and the changes in the properties such as MCE and resistivity, in the substituted compounds. Several studies have also revealed that the variation of adiabatic temperature change is identical to that of the entropy change. The maximum values of $\Delta T_{ad}$ obtained in some of the compounds are also listed in Table 1.

The effect of substitutions on the magnetic properties and MCE of these compounds could be understood by calculating the temperature variation of the Landau coefficients. It is well known that the magnetic free energy, *F(M,T)*, in general can be expressed as Landau expansion in the magnetization (see equation 12) and the temperature and magnetic field dependence of *F(M,T)* determines the nature of magnetic transition. The Landau coefficients can be calculated using the equation of state, given by:

$$H = AM + BM^3 + CM^5 \qquad (15)$$

It may be noted from equation 15 that the magnetization isotherms obtained at various temperatures may allow one to determine the temperature variation of the Landau coefficients. It is well known that the temperature dependence of the Landau coefficients may be utilized to distinguish between the first and second order transitions of magnetic materials [23, 34, 54]. It may be noted from equation 15 that the coefficient *A* corresponds to inverse of susceptibility and therefore a minimum in the temperature dependence of *A* is expected at the ordering temperature. The materials with negative *A* are known to exhibit a second order transition whereas those having positive *A* but negative *B* exhibit first order transition. In the following, the Landau analysis of the magnetization data is demonstrated by taking the example of Si substitution for Co in ErCo$_2$.

The temperature variation of Landau coefficient (*B*) obtained for Er(Co$_{1-x}$Si$_x$)$_2$ compounds with x=0 and 0.5 is shown in Fig. 13. It may be noted from the figure that for both the compounds the sign of *B* near T$_C$ is negative and that its magnitude decreases with increase in temperature. Therefore the temperature variation of *B* in both the compounds indicates the presence of FOT at T$_C$. It may also be seen that the magnitude of *B* in Er(Co$_{0.95}$Si$_{0.05}$)$_2$ at temperatures well below T$_C$ is lower than that of ErCo$_2$. This could be ascribed to the increase in the spin fluctuation contribution on Si substitution. The effect of substitutions, as mentioned earlier, is to suppress the IEM and hence the *B* values.

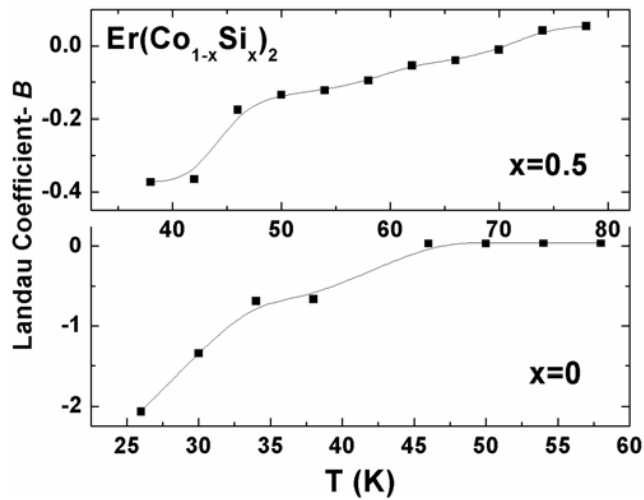

Fig. 13 Temperature (T) dependence of the Landau coefficient $B$ in $Er(Co_{1-x}Si_x)_2$ compounds with x=0, 0.05. [$B$ values have been calculated in c.g.s. units]. [Data taken from Ref. 42.]

Table 2 Isothermal entropy change in $Er(Co_{1-x}Si_x)_2$ compounds under various pressures.

| ErCo$_2$ | | Er(Co$_{0.975}$Si$_{0.025}$)$_2$ | | Er(Co$_{0.95}$Si$_{0.05}$)$_2$ | |
|---|---|---|---|---|---|
| P (kbar) | $\Delta S_M^{max}$ (Jkg$^{-1}$K$^{-1}$) | P (kbar) | $\Delta S_M^{max}$ (Jkg$^{-1}$K$^{-1}$) | P (kbar) | $\Delta S_M^{max}$ (Jkg$^{-1}$K$^{-1}$) |
| Ambient* | 33 | Ambient* | 27.4 | Ambient* | 22.7 |
| 4.7 | 32.8 | 2.7 | 28.8 | 3.3 | 24.8 |
| 7.4 | 32.5 | 5.8 | 29.9 | 9 | 26.6 |

* Ambient pressure = 1 bar.

### 4.1. Effect of pressure on the magnetocaloric effect

As mentioned in the earlier sections, there are several MCE studies in literature regarding the substituted $RCo_2$ compounds. However, the studies of MCE under applied pressure are very rare in these compounds. The only work available to this effect is on $ErCo_2$ systems. Fig. 14 shows the MCE variation in $ErCo_2$ and $Er(Co_{0.95}Si_{0.05})_2$ compounds under various pressures. The pressure dependence of $\Delta S_M^{max}$ of both these compounds is given in Table 2. It can be seen from Fig. 14 (also from Table 2) that, with increase in pressure, the peak in the $\Delta S_M$ vs T plot moves towards low temperatures in both the compounds. However, in the parent compound, the value of $\Delta S_M^{max}$ almost remains insensitive to pressure while in the substituted compounds, it is found to increase. The insensitiveness of MCE on applied pressure in the case of $ErCo_2$ is due to the fact that the strength of IEM has diminished only nominally even at a pressure of about 7.7 kbar. Hauser *et al.* [55] have indeed reported that the discontinuity (at $T_C$) in the magnetic contribution to the electrical resistivity in ErCo$_2$ decreases to about 60 % as the pressure is increased from 1 bar to ~16 kbar, which is attributed to the reduction in the strength of IEM. In view of this, it is reasonable to assume that for a pressure of 7.7 kbar, the reduction in the IEM strength is not very much and therefore would contribute only to a nominal reduction in $\Delta S_M^{max}$. However, the reduction in $T_C$ brought about by pressure would try to enhance $\Delta S_M^{max}$ due to the reduction in the thermal spin fluctuations. Therefore, it is quite possible that the reduction in MCE caused by the weakening of IEM is just compensated by the increase in MCE arising out of the reduction in $T_C$. Therefore, the insensitiveness of MCE on pressure seen in the case of ErCo$_2$ is consistent with the observations made by Hauser et al. [55].

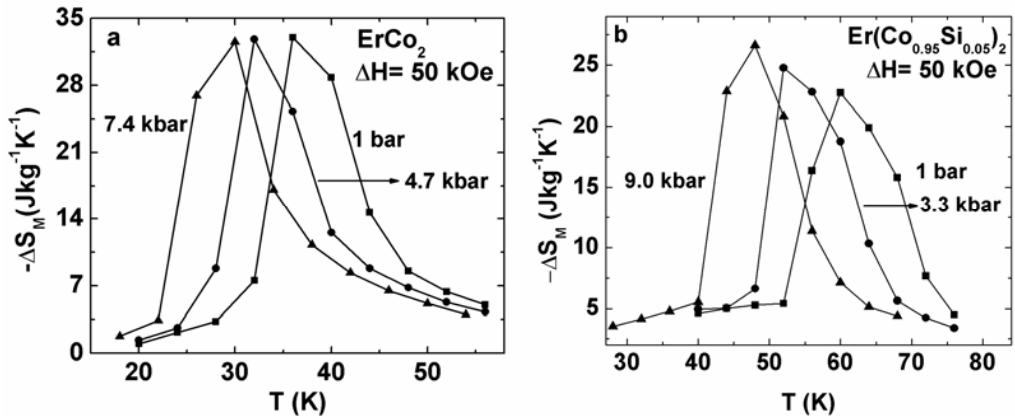

Fig. 14 Temperature dependence of isothermal entropy change ($\Delta S_M$) in Er(Co$_{1-x}$Si$_x$)$_2$ compounds obtained for $\Delta H$= 50 kOe, under various applied pressures (P). [Figure taken from Ref. 42]

On the other hand, the pressure dependence of $\Delta S_M^{max}$ in the case of Si substituted compounds is quite considerable. The increase seen in $\Delta S_M^{max}$ may be attributed to the increase in the strength of IEM, which results from the reduction of spin fluctuations related to the negative magnetovolume effect. As in the case of substitutions, Landau analysis can be used to study the changes in the nature of transitions and MCE, with increase in pressure as well. The temperature variation of Landau coefficient (*B*) obtained for Er(Co$_{1-x}$Si$_x$)$_2$ compounds with x=0 and 0.5 for different pressures is shown in Fig. 15. There is no change in the low temperature value of B in the case of ErCo$_2$, while there is a considerable increase in Er(Co$_{0.95}$Si$_{0.05}$)$_2$, with increase of pressure. This implies that the strength of IEM increases considerably with pressure in the Si substituted compounds whereas there is no significant change in ErCo$_2$. It may also be noticed from this figure that the difference between the low temperature *B* values of ErCo$_2$ and the Si-substituted compounds almost vanishes with increase in pressure. The effect of applied pressure (in the substituted compounds) is to compete with the positive magnetovolume effect caused by substitutions and to reduce it. Consequently, the magnetic nature of the Co sublattice in the Si-substituted compounds is more or less restored to that of ErCo$_2$, by the applied pressure. This is exactly seen in Fig.15 which shows that at high pressures, the low temperature *B* value of Er(Co$_{0.95}$Si$_{0.05}$)$_2$ is almost equal to that of ErCo$_2$.

At this point, it is of importance to note that the variation of Landau coefficient (B) with pressure in the case of parent compounds showed that it is insensitive to pressure whereas in the case of Si substituted compounds, it showed an increase in the magnitude. Therefore, the variations in the magnitude of *B* with pressure are consistent with the MCE variation, which implies that there is a strong correlation between the *B* value and the MCE in RCo$_2$-based IEM systems. In this context, it is of relevance to mention that Yamada *et al.* [56] have indeed shown that the MCE in IEM systems is primarily governed by the magnitude of *B*. Fujita *et al.* have also reported a similar dependence of MCE on *B* in La(Fe,Si)$_{13}$ compounds which also is a well known IEM system [57]. Based on the magnetization behaviour, it is expected that other nonmagnetic substitutions in RCo$_2$ [R=Er, Ho and Dy] compounds would also result in similar correlation between MCE and the Landau coefficient, *B*.

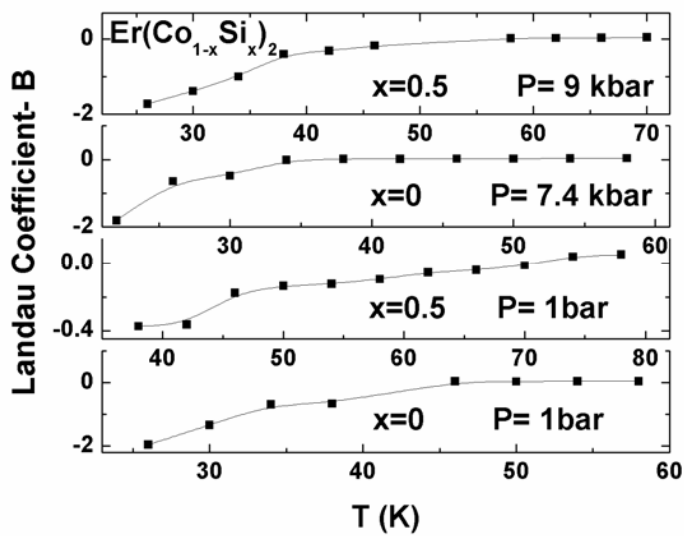

fig. 15 Temperature (T) dependence of the Landau coefficient *B*, obtained under various external pressures (P), in Er(Co$_{1-x}$Si$_x$)$_2$ compounds with x=0, 0.05. [*B* values have been calculated in cgs units]. [Figure taken from Ref.42]

## 5. SUMMARY AND CONCLUSIONS

In conclusion, the magnetic and magnetocaloric studies on the RCo$_2$-based compounds show that ErCo$_2$, HoCo$_2$ and DyCo$_2$ exhibit first order magnetic transition and large MCE. The nonmagnetic nature of Co sublattice in these compounds causes IEM at temperatures close to the ordering temperature, which leads to first order transition. Positive magnetovolume effect resulting from the substitution of nonmagnetic elements for Co is found to suppress IEM and FOT, due to the enhanced spin fluctuations. The spin fluctuations arising from the local moment formation are found to be detrimental to MCE in the substituted compounds. On the other hand, application of pressure destabilizes the Co sublattice magnetization because of the negative magnetovolume effect and therefore, the ordering temperature decreases. The effect of pressure on the MCE is clearly seen in the substituted compounds in which, the negative magnetovolume effect competes with the positive magnetovolume effect resulting from the substitutions. The thermodynamic analysis based on the Landau's theory reveals that the MCE variations follow the variations seen in the magnitude of the coefficient B of the Landau free energy expression. The variations seen in the order of magnetic transition and the MCE values seem to support the recent model proposed by Khmelevskyi and Mohn for the occurrence of IEM in $RCo_2$ compounds.

## ACKNOWLEDGEMENTS

The authors would also like to express their gratitude to R. Nirmala, Pramod Kumar, Devendra Buddhikot for their help in various measurements. One of the authors (KGS) would like to thank DST and BRNS (DAE), Govt. of India for financial support to carry out some of the work mentioned in this report.

## REFERENCES

[1] Jiles, D. C. 2003, Acta Materialia 51, 5907.
[2] Pecharsky, V. K. and Gschneidner, K. A. Jr. 1999, J. Mag. Mag. Mater. 200, 44.
[3] Bruck, E. 2005, J. Phys. D: Appl. Phys. 38, R381.
[4] Warburg, E. 1881, Ann. Phys. 13, 141.
[5] Debye, P. 1926, Ann. Phys. 81, 1154.
[6] Giauque, W. F. 1927, J. Amer. Chem. Soc. 49, 1864.
[7] Gschneidner, K. A. Jr. 2002, J. Alloys Comp. 344, 356.
[8] Tishin, A. M., Gschneidner, K. A. Jr. and Pecharsky, V. K. 1999, Phys. Rev. B 59, 503.
[9] Foldeaki, M., Chahine, R. and Bose, T. K. 1995, J. Appl. Phys. 77, 3528.
[10] Pecharsky, V. K. and Gschneidner, K. A. Jr. 1999, J. Appl. Phys. 86, 565.
[11] Zemansky, M. W. 1981, Heat and Thermodynamics, McGraw-Hill (6th ed.), New York.
[12] Pecharsky, V. K. and Gschneidner, K. A. Jr. 1996, Adv. Cryog. Engg., 42, 423.
[13] Giauque, W. F. and MacDougall, I. D. P. 1933, Phys. Rev. B 43, 768.
[14] Ishimoto, H., Nishida, N., Furubayashi, T., Shinohara, M., Takano, Y., Miura, Y. and Ono, K. 1984, J. Low Temp. Phys. 55, 17.
[15] Levitin, R. Z., Snegirev, V. V., Kopylov, A. V., Lagutin, A.S. and Gerber, A. 1997, J. Mag. Mag. Mater. 170, 223.
[16] Gschneidner, K. A. Jr., Pecharsky, V. K. and Malik, S. K. 1996, Adv. Cryog. Engg. 42, 475.
[17] Singh, N. K., Agarwal, S., Suresh, K. G., Nirmala, R., Nigam, A. K. and Malik, S. K. 2005, Phys. Rev. B 72, 014452.
[18] Takeya, H., Pechasky, V. K., Gschneidner, K. A. Jr. and Moorman, J. O. 1994, Appl. Phys. Lett. 64, 2739; Singh, N. K., Suresh, K. G., Nirmala, R., Nigam, A. K. and Malik, S. K. 2006, J. Appl. Phys. 99, 08K904.
[19] Pecharsky, V. K. and Gschneidner, K. A. Jr. 1997, Appl. Phys. Lett. 70, 3299.
[20] Pecharsky, V. K. and Gschneidner, K. A. Jr. 2001, J. Appl. Phys. 90, 4614.
[21] Wada, H. and Tanabe, Y. 2001, Appl. Phys. Lett. 79, 3302.

[22] Tegus, O., Bruck, E., Buschow, K. H. J. and Boer, F. R. de 2002, Nature **415**, (2002) 150.
[23] Gratz, E. and Markosyan, A. S. 2001, J. Phys. Condens. Matter 13, R385.
[24] Wada, H., Tanabe, Y., Shiga, M., Sugawara, H. and Sato. H. 2001, J. Alloys Compd 316, 245.
[25] Duc, N. H., Kim, D. T. Anh and Brommer, P. E. 2002, Physica B 319, 1.
[26] Prokleska, J., Vejpravova, J., Vasylyev, D. and Sechovsky, V. 2004, J. Alloys Compd 383, 122.
[27] Singh, N. K., Suresh, K. G., Nigam, A. K. and Malik, S. K. 2005, J. Appl. Phys. **97** 10A301.
[28] Tohei, T. and Wada, H. 2004, J. Mag. Mag. Mater. 280, 101.
[29] Wang, D., Tang, S., Liu, H., Zhong, W. and Du, Y. 2003, Mat. Letts. 57, 3884.
[30] Han, Z., Hua, Z., Wang, D., Zhang, C., Gu, B. and Du, Y. 2006, J. Mag. Mag. Mater. 302, 109.
[31] Buschow, K. H. J. 1980, Ferromagnetic materials, Wohlfarth, E. P. (Ed.) Vol. 1, North Holland, Amsterdam.
[32] Duc, N. H., Hien, T. D., Brommer, P. E. and Franse, J. J. M. 1992, J. Mag. Mag. Mater. 104-107, 1252.
[33] Hauser, R., Bauer, E., Gratz, E., Müller, H., Rotter, M., Michor, H., Hilscher, G., Markosyan, A. S., Kamishima, K. and Goto, T. 2000, Phys. Rev B 61, 1198; Gignoux, D. and Schmitt, D. 1991, J. Mag. Mag. Mater. 100, 99.
[34] Fujita, A., Fujieda, S., Hasegawa, Y. and Fukamichi, K. 2003, Phys. Rev. B 67, 104416.
[35] Singh, N.K., Suresh, K. G. and Nigam A. K. 2003, Solid State Commun. 127, 373.
[36] Wohlfarth, E. P. and Rhodes, P. 1962, Philos. Magm **7**, 1817.
[37] Ohta, M., Fukamichi, K., Fujita, A., Saito, H. And Goto, T. 2005, J. Alloy Compd. 394, 43.
[38] Duc, N. H. and Brommer, P. E. 1999, Handbook on Magnetic Materials, Buschow, K. H. J. (Ed.) Vol. 12, North Holland, Amsterdam.
[39] Bloch, D., Ewards, D. M., Shimizu, M. and Voiron, J. 1975, J. Phys. F. Met. Phys. 5, 1217.
[40] Inoue, J. and Shimizu, M. 1982, J. Phys. F. Met. Phys. 12, 1811.
[41] Khmelevskyi, S. and Mohn, P. 2000, J. Phys. Condens. Matter 12, 9453.
[42] Singh, N. K., Kumar, P., Suresh, K. G., Nigam, A. K., Coelho, A. A. And Gama, S. 2006, cond-mat/0608191.
[43] Vasylyev, D., Prokleska, J., Sebek, J. and Sechovsky, V. 2005, J. Alloys Compd. 394, 96.
[44] Duc, N. H., Hien, T. D., Levitin, R. Z., Markosyan, A. S., Brommer, P. E. and Franse J. J. M. 1992, J. Alloys Compd. 176, 232.
[45] Prokleska, J., Vejpravova, J., Vasylyev, D., Danis, S. and Sechovsky, V. 2005, J. Mag. Mag. Mater. 290-291, 676.
[46] Baranov, N. V. and Pirogov, A. N. 1993, J. Alloys Compd. 202, 17.
[47] Ouyang, Z. W., Rao, G. H., Yang, H. F., Liu, W. F., Liu, G. Y., Feng, X. M. and Liang J. K. 2004, Physica B 344, 436.
[48] Podlesnyak, A., Teplykh, A., Teplykh, P., Krivoshchekov, A. and Sadykov, R. 2004, Physica B 350, E143.
[49] Syshchenko, O., Fujita, T., Sechovsky, V., Divis, M. and Fujii, H. 2001, Phys. Rev. B 63, 054433.

[50] Yamada, H. 1995, J. Mag. Mag. Mater. 139, 162.
[51] Gschneidner, K. A. Jr., Pecharsky, V. K. and Tsokol, A. O. 2005, Rep. Prog. Phys. 68, 1479.
[52] Singh, N. K., Kumar, P. , Suresh, K. G. and Nigam, A.K. (Unpublished)
[53] Duc, N. H., Sechovsky, V., Hung, D. T. and Kim-Ngan, N. H. 1992, Physica B 179, 111.
[54] Yamada, H. 1993, Phys. Rev. B. 47, 11211.
[55] Hauser, R., Bauer, E. and Gratz, E. 1998, Phys. Rev. B 57, 2904.
[56] Yamada, H. and Goto, T. 2003, Phys. Rev. B. 68, 184417.
[57] Fujita, A. and Fukamichi, K. 2005, IEEE Trans. Magn. 41, 3490.